\documentstyle[aps,floats,twocolumn,prd,psfig]{revtex}

\newcommand{\ApJ}{Astrophys. J.} 
\newcommand{\PRL}{Phys. Rev. Lett.}
\newcommand{\PRD}{Phys. Rev. D}
\newcommand{\MNRAS}{Mon. Not. Roy. Astron. Soc.}

\def\sun{\hbox{$\odot$}}

\newlength{\tskip}
\newlength{\colwidth}

\newcommand{\beq}{\begin{equation}}
\newcommand{\eeq}{\end{equation}}
\newcommand{\beqa}{\begin{eqnarray}}
\newcommand{\eeqa}{\end{eqnarray}}

\long\def\comment#1{}

\begin{document}   

\twocolumn[\hsize\textwidth\columnwidth\hsize\csname @twocolumnfalse\endcsname
\title{CMB Polarization towards Clusters as a Probe of the Integrated Sachs-Wolfe Effect}
\author{Asantha Cooray$^{1}$ and Daniel Baumann$^{1,2}$}
\address{$^1$California Institute of Technology, Mail Code
130-33, Pasadena, CA 91125\\
$^2$University of Cambridge, Cambridge, CB3 OHE\\
E-mail: (asante,db275)@tapir.caltech.edu}

\maketitle
\begin{abstract}
The scattering of temperature anisotropy quadrupole by
free electrons in galaxy clusters leads to a now well-known 
polarization signal in the cosmic microwave background (CMB) fluctuations. 
Using multi-frequency polarization data, one can extract the
temperature quadrupole and separate it from the contaminant 
polarization associated with the kinematic quadrupole due to transverse motion of clusters.
At low redshifts, the temperature quadrupole contains
a significant contribution from the integrated Sachs-Wolfe effect (ISW) 
associated with the growth of density fluctuations. 
Using polarization data from a sample of clusters over a wide range in
redshift, one can statistically establish the presence of the ISW effect 
and determine the redshift dependence of the ISW contribution to the rms quadrupole.
Given the strong dependence of the ISW effect on the background cosmology, 
the cluster polarization can eventually be used as a 
probe of the dark energy.
\end{abstract}

\hfill
]

\section{Introduction}
Linear polarization of the cosmic microwave background (CMB) is generated
through rescattering of the temperature quadrupole~\cite{kaiser}.
 At the last scattering surface, this
leads to the primordial polarization contribution~\cite{HuWhi97}, 
while at lower 
redshifts, the 
temperature 
quadrupole gets scattered in regions containing free electrons to produce a 
secondary polarization signal. One such possibility arises from reionization 
which generates a large angular scale polarization  contribution corresponding to the horizon
at the surface of reionization~\cite{Zal97}. 
At much smaller angular scales corresponding to few arcminutes
and below, secondary polarization is generated when the quadrupole is
scattered by free electrons in galaxy clusters~\cite{SunZel80}.

There are several mechanisms to generate a polarization 
signal in clusters. However, 
scattering of the primordial temperature anisotropy quadrupole is expected to
 dominate
the total contribution~\cite{SazSun99,Challinor}. A second effect
arises from the 
kinematic quadrupole 
due to the transverse motion of the cluster. 
This contribution, fortunately, has a different
frequency dependence than the contribution associated with the scattering of
 the primordial quadrupole. 

Following initial suggestions on the cluster polarization contributions \cite{SunZel80}, 
recent studies have focused on detailed 
aspects related to spectral dependences \cite{SazSun99,Challinor}, 
the statistical signature such as the angular power
spectra of polarization \cite{Hu99},
 and the use of cluster polarization as a mechanism to
improve the local determination of the temperature quadrupole \cite{KamLoe97}.

Here, we suggest an extension to the discussion in Ref.~\cite{KamLoe97}.
The basic idea is to use galaxy clusters as a tracer of the
temperature anisotropy quadrupole and to statistically detect its rms value.
Our discussion involves the measurement of quadrupole as a function of redshift
or redshift bins based on a cluster catalog. We improve the previous discussion  in Ref.~\cite{KamLoe97} 
by considering the contaminant kinematic quadrupole contribution and
introducing the use of multi-frequency observations.
We make a quantitative prediction for the prospects of measuring the
redshift evolution of the temperature quadrupole with future CMB
polarization experiments.

The statistical reconstruction of the rms temperature quadrupole, as a
 function of redshift, is useful for several purposes.  
At low redshifts, the quadrupole
contains information related to growth of structures via the integrated Sachs-Wolfe (ISW) effect 
\cite{SacWol67}.
The presence of the ISW effect, however, cannot be established from the 
temperature anisotropy
measurements alone, due to the dominant Sachs-Wolfe (SW) 
contribution and the large sample variance
associated with low multipoles. 

Since the ISW contribution mainly arises from low redshifts, one expects large scale CMB contributions to be correlated with maps of 
the large scale structure \cite{CriTur96}. 
This correlation, however, is hard to establish observationally
both due to large cosmic variance and the lack of adequate signal-to-noise
all-sky maps of tracer fields \cite{Coo02}.  
One therefore needs a probe of the ISW effect that is not subject to a large sample
variance and we propose the use of cluster polarization for this purpose.
Using the polarization signal towards a sample of clusters widely distributed over the whole sky,
one can establish the presence of the ISW effect reliably.  Additionally, since 
the ISW contribution is strongly sensitive to the background
cosmology, the redshift evolution of  the rms quadrupole 
can be used as a probe of the dark energy.

The paper is organized as follows.  In \S~\ref{sec:pol}, we
introduce polarization signals towards clusters associated with the
scattering of the primordial CMB temperature quadrupole and with
the kinematic quadrupole generated by transverse cluster motions and study the use of
multi-frequency polarization observations as a way 
to separate these two contributions.  We will then consider the
reconstruction of the rms temperature quadrupole as a function of
redshift using polarization observations towards 
a sample of galaxy clusters. We discuss our results in \S~\ref{sec:results}. 

\section{Polarization towards Clusters}
\label{sec:pol}

CMB polarization towards clusters is generated when the
incident radiation has a nonzero 
quadrupole moment. The two dominant origins for this quadrupole
moment are: (a) A quadrupole from primordial CMB fluctuations, and
(b) a quadrupole from the quadratic term in the Doppler shift if
the scattering gas has a transverse velocity \cite{SazSun99,Challinor,Hu99}.

Towards a sufficiently large sample of galaxy clusters, we can write the total
rms polarization contribution as~\cite{SazSun99}
\begin{eqnarray}
P^{\rm Tot} &=& P^{\rm Prim}+P^{\rm Kin}\nonumber \\
P^{\rm Prim}&=&\frac{\sqrt{6}}{10} \langle \tau \rangle \frac{Q^{\rm rms}(z)}{T_{\rm CMB}}\nonumber \\
P^{\rm Kin}&=&\frac{1}{10} g(x) \langle \tau \rangle \langle \beta_t^2\rangle \, , \nonumber \\
\label{eqn:ptot}
\end{eqnarray}
where $\tau$ is the optical depth to scattering in each cluster: $\tau = \sigma_T\int dy  n_e(y) $.  For an individual cluster,  the optical depth can be constrained through 
the temperature fluctuation associated with the Sunyaev-Zel'dovich \cite{SunZel80} effect. 
Since we are averaging over large samples of clusters, in Eq.~\ref{eqn:ptot},
we have considered the sample averaged optical depth, $\langle \tau \rangle$.
$\beta_t=v_t/c$ gives the transverse component of the cluster
velocity and $g(x)$, with $x=h\nu/k_B T_{\rm CMB}$, is the frequency dependence
of the kinematic effect (discussed below).

Note that the rms quadrupole of temperature anisotropies, as a function of
redshift, is related to the $l=2$ contribution to the CMB power spectrum as measured by an observer at that redshift: 
$Q_{\rm rms}^2(z) = 5C_2(z)/4\pi$. Large angular scale observations,
such as from COBE, constrain the variance of the temperature
quadrupole today to be $C_2(z=0)=(27.5 \pm 2.4\mu{\rm K})^2$, assuming a 
power-spectrum tilt of unity \cite{Ben94}. 
The quadrupole is highly uncertain due to
the fact that one is limited to only five independent samples.
Also, these observations do not allow a
determination of the rms quadrupole at redshifts other than today.
However, measuring the CMB polarization towards galaxy clusters
potentially allows to probe the quadrupole moment at the cluster
position~\cite{KamLoe97}.

We can calculate the expected redshift evolution of the quadrupole based on the
two sources of contributions, one at the surface of last scattering due to
SW effect and another at late times due to the 
ISW effect associated with a time evolving potential \cite{SacWol67}. We write these two contributions to the power spectrum, projected to a redshift $z$, as
\begin{eqnarray}
C_{l}(z) &=& C_l^{\rm SW}(z)+C_l^{\rm ISW}(z)\, , \nonumber \\
C_l^{\rm SW}(z) &=& \frac{\Omega_m^2 H_0^4}{2 \pi a^2(r_{\rm rec})}
\int_0^{\infty} \frac{dk}{k^2} P(k,r_{\rm rec}) j_l^2[k(r_{\rm rec}-r_z)] \,,\nonumber \\
C_l^{\rm ISW}(z) &=&  \frac{18 \Omega_m^2 H_0^4}{\pi} \int_0^{\infty} 
\frac{dk} {k^2} P(k,0) \nonumber \\
&\times&\left[\int_{r_z}^{r_{rec}} dr'_z \frac{d}{dr'_z}\left(\frac{G}{a}\right)\ j_l(k(r'_z-r_z))\right]^2 \, .
\label{eqn:cl}
\end{eqnarray}
Here, $r_z$ is the radial comoving distance out to redshift  $z$,
$r_{\rm rec}$ is the distance to last scattering ($z=1100$), and
$G(r_z)$ is the growth rate of linear density fluctuations with
$\delta^{\rm lin}(k,r_z)=G(r_z)\delta^{\rm lin}(k,0)$ \cite{Pee80}. Here, 
$\delta^{\rm lin}(k,r_z)$ is the fluctuation in the linear density field
and $P(k,r_z)$, in Eq.~\ref{eqn:cl}, is its power spectrum. 

In addition to scattering via the temperature anisotropy quadrupole,
the kinematic quadrupole associated with the transverse motion also
produces a polarization signal towards galaxy clusters
\cite{SunZel80,SazSun99,Challinor}. To understand how this
polarization contribution is generated, first, consider electrons
moving with peculiar velocity, $\beta=v/c$, relative to the rest frame defined by the CMB.
The CMB spectral intensity in the mean electron rest frame is Doppler shifted
\begin{equation}
I_{\nu} = C \frac{x^3}{e^{x \gamma(1+\beta \mu)}-1}\, ,
\end{equation}
where $x \equiv h \nu/ k_B T_{CMB}$ and 
$\mu$ is the cosine of the angle between 
the cluster velocity and the direction of the incident CMB photon.
When expanded in Legendre polynomials, the intensity distribution is
\begin{equation}
I_{\nu} = C \frac{x^3}{e^x-1} \left[ I_0 + I_1 \mu + \frac{e^x(e^x+1)}{2(e^x-1)^2}x^2 \beta^2 \left(\mu^2-\frac{1}{3}\right) + \dots \right] \, 
\end{equation}
and contains the necessary quadrupole under which scattering generates
a polarization contribution with Stokes-parameters given by
$Q(\mu')=1/10g(x)\beta^2 (1-\mu')^2=1/10g(x)\beta_t^2$, $U(\mu')=0$ where $\mu'$ is now the angle between
cluster velocity and the line of sight.

We can write the spectral dependence associated with  the
kinematic quadrupole, relative to CMB thermal temperature, as
\begin{equation}\label{equ:QuadKin}
g(x) =  \frac{x}{2}\coth \left(\frac{x}{2}\right)\, .
\label{eqn:kinI}
\end{equation}
Note that the spectral dependence of the kinematic quadrupole
contribution gives a potential
method to separate the two contributions. This is similar to component
separation suggestions in the literature as applied to temperature 
observations, such as the separation of the SZ thermal effect from
dominant primordial fluctuations \cite{Cooetal00}.

\begin{figure}[!h]
\centerline{\psfig{file=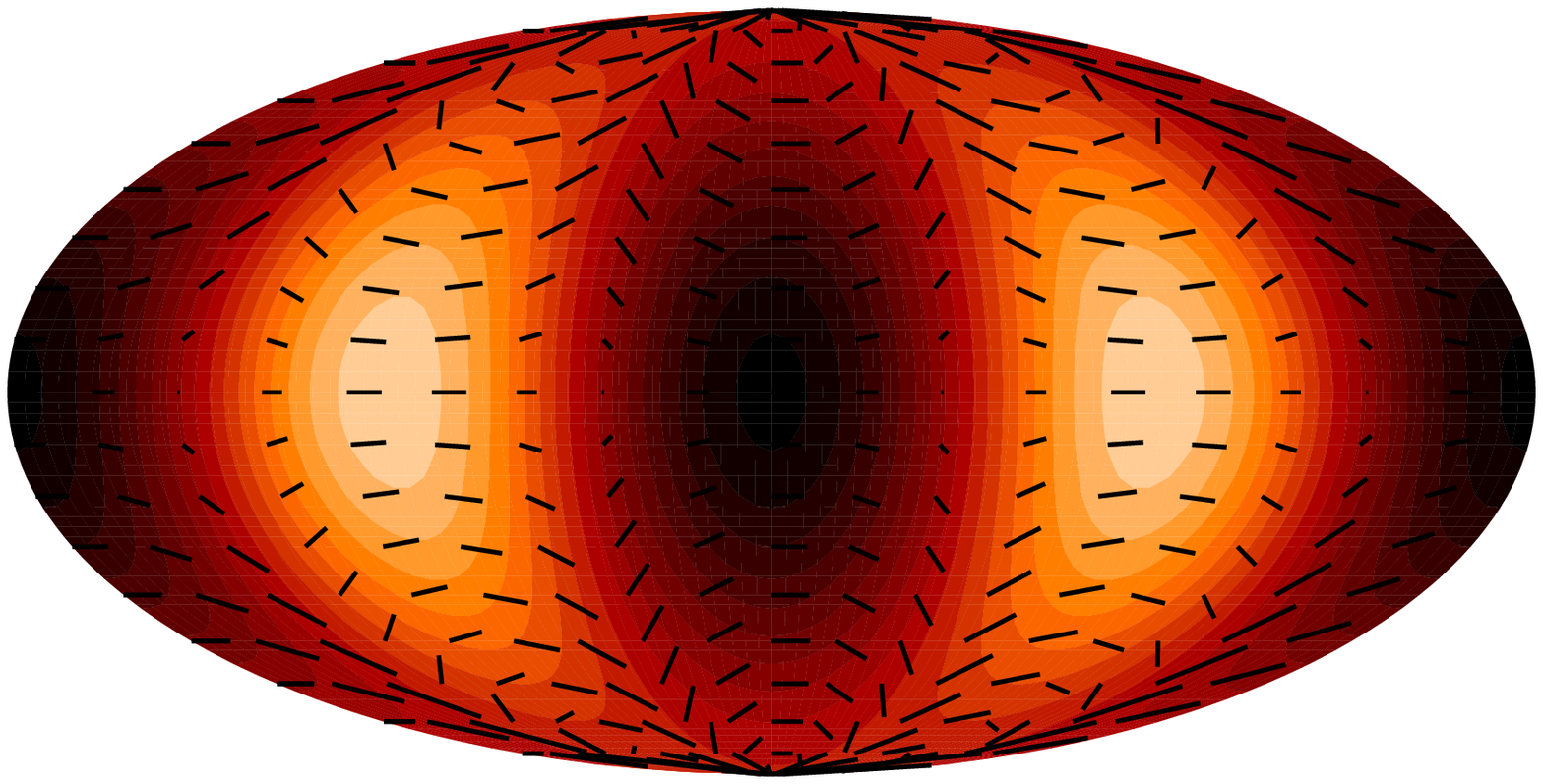,width=3.1in}}
\centerline{\psfig{file=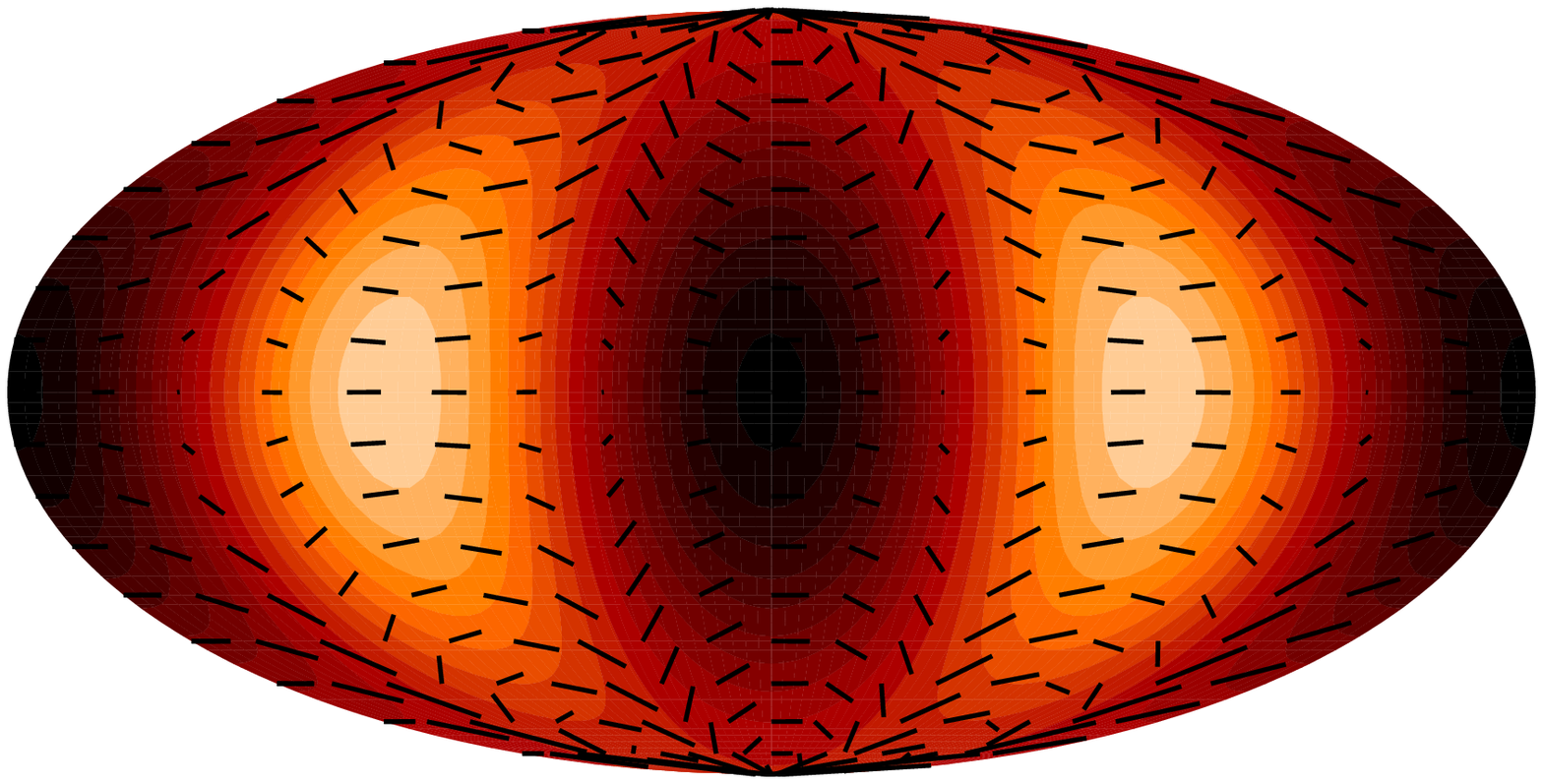,width=3.1in}}
\centerline{\psfig{file=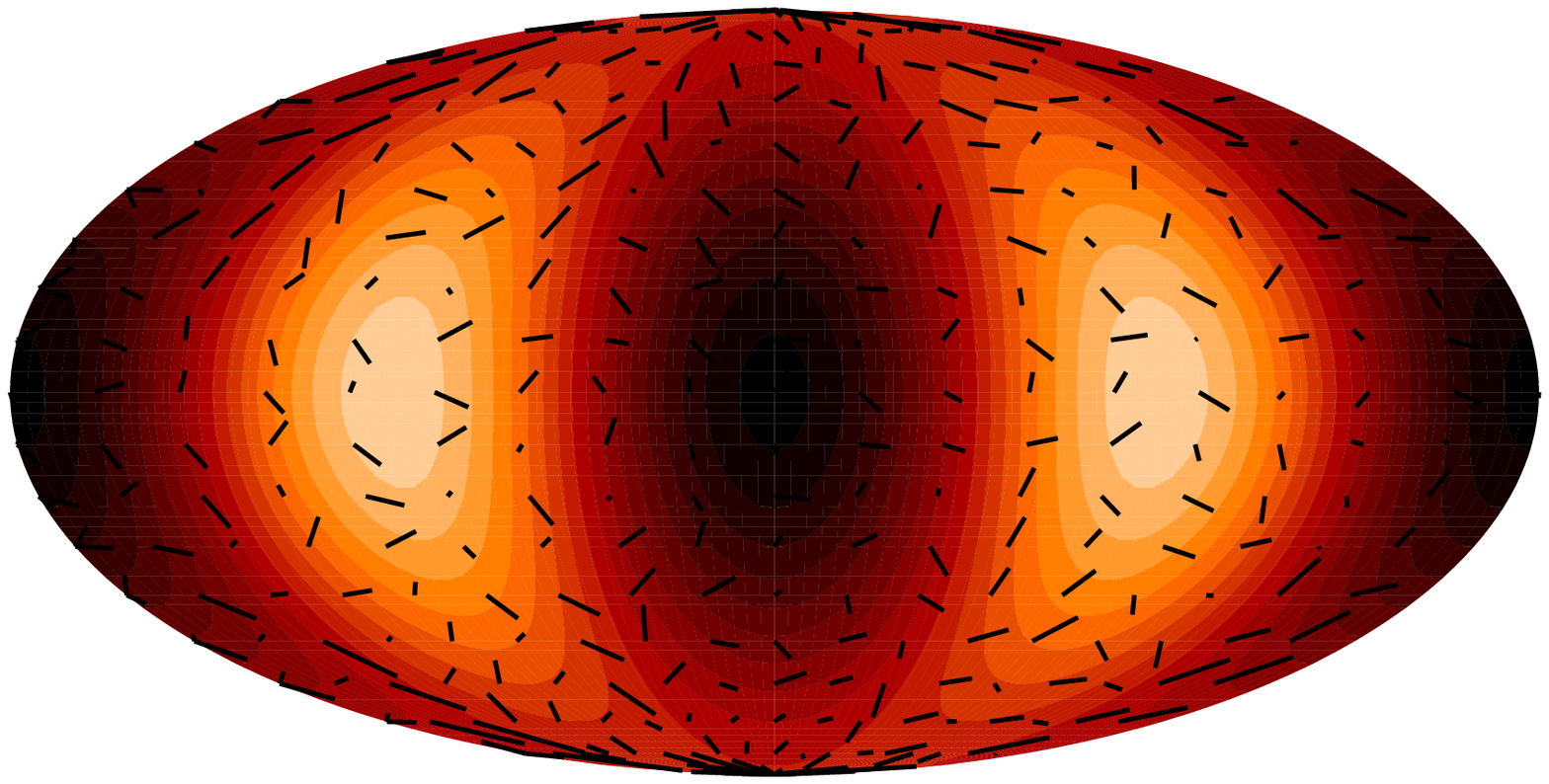,width=3.1in}}
\caption{The CMB polarization due to galaxy clusters. 
The polarization vectors give a representation of the expected
polarization from a cluster at the corresponding location in the sky.
The scale is such that the maximum length of the line corresponds to a polarization of 4.9 $\tau$ $\mu$K.}
The background color
represents the temperature quadrupole. The top plot shows
the resulting polarization contribution due to scattering of the 
primordial  temperature quadrupole alone. 
The  middle map show the total polarization contribution (primordial
and kinematic) at 600 GHz; the total
is clearly dominated by the primordial contribution.
To highlight the kinematic polarization, in the bottom map,
we arbitrarily increase its amplitude by a factor of 100.
Note that even at high frequencies where the kinematic quadrupole is
 increased due to its spectral dependence,
given in Eq.~\ref{equ:QuadKin}, the
primordial polarization still dominates the total contribution.
\label{fig:prim}
\end{figure}

\subsection{Frequency Separation}

Here, we will consider separation based on real space data
instead of the multipolar analysis considered for
the temperature \cite{Tegetal99}.
We will briefly discuss the separation
problem in the presence of noise. We refer the reader to
Ref. \cite{Dod97} for further details. In the present case, the problem involves the extraction of
the polarization associated with the primordial quadrupole, 
with the kinematic quadrupole considered as a source of confusion.
We can write the total polarization towards a sample of
clusters within a redshift bin as
\begin{equation}
{\bf P}= {\bf Q} + {\bf K} + {\bf N}\, ,
\end{equation}
where ${\bf Q }$ is the contribution due to the
primordial quadrupole, ${\bf K}$ is the contribution due to 
 the kinematic quadrupole and ${\bf N}$ characterizes the instrumental noise.
The two-dimensional  vector field 
associated with polarization can be decomposed into
two scalar quantities involving two orthogonal measurements, which we
will simply denote as $(P_x,P_y)$. These orthogonal quantities
can be related to the usual Stokes parameters involving Q and U or the gradient and
curl representation of the spin-2 field involving E and B-modes \cite{EB}.

We assume Gaussian noise with
\begin{eqnarray}
\left<N_x\right> &=& 0\, , \nonumber \\
\left<N_x^a N_x^b\right> &=& C^x_{ab}\, ,
\end{eqnarray}
and similarly for the $y$-component.
Here, $a,b=1,...,N_{\rm ch}$ labels the different frequency channels of the 
experiment.

In Fig.~\ref{fig:prim}, we illustrate the frequency dependence of the
primordial and kinematic quadrupole contributions using all-sky
maps of the expected polarization. 
In these plots, each polarization vector, shown as lines, 
should be considered as a representation of the polarization towards a 
cluster at that location. The polarization pattern due to
temperature quadrupole scattering is uniform and traces the underlying
temperature quadrupole distribution. 
In describing the kinematic quadrupole we assume, for illustration purposes,
a transverse velocity field with $\langle \beta_t \rangle \sim 
10^{-3}$ corresponding to a velocity of 300 km sec$^{-1}$.
The polarization contribution due to
the kinematic quadrupole, however, is random due to that fact that 
transverse velocities are uncorrelated; the
correlation length of the velocity field of order 60 Mpc correlates
velocities within regions of 1$^{\circ}$ when projected to a redshift of
order unity. Such a correlation scale is much smaller than the
all-sky maps we have considered in Fig.~\ref{fig:prim}.

As shown in Fig.~1, even at high frequencies where 
the kinematic quadrupole is increased due to the spectral dependence,
the primordial polarization still dominates the total contribution.
This is also true when allowing for the fact that we may have underestimated 
the transverse velocity field; while the maximum primordial quadrupole
related polarization remains at $4.9\tau$ $\mu$K, the 
polarization associated with the kinematic quadrupole 
scales as $0.27 g(x) (\beta_t/0.001)^2 \tau$ $\mu$K.  To get a comparable
contribution at frequencies of order few 100 GHz with $g(x) \sim$ few, 
the transverse velocity field should involve magnitudes of order 
1000 km sec$^{-1}$; such a high velocity is unlikely to be present
in the large scale structure.
The fact that primordial polarization dominates over the kinematic
contribution even at high frequencies 
is useful for the present study since the 
reconstruction of the primordial quadrupole
is not likely to be confusion limited. 

To find the best-fit estimator ${\cal P}^0$ associated with 
${\bf Q}$, and its variance,  we calculate, for say one of the components of
polarization,
\begin{equation}
\frac{\delta}{\delta {\cal P}_x^i}\left<\left(P_x^a-\sum_{i=0}^1 g^i {\cal P}_x^
i\right)^2\right> =0\, .
\end{equation}
Here, $i=0$ denotes the ${\bf Q}$ contribution and  $i=1$ denotes the ${\bf K}$ 
contribution.  $g^i$ defines the spectral dependence of both effects; 
by definition $g^0=1$  as we consider variation of the kinematic quadrupole 
contribution  relative to the thermal CMB. $g^1$ is given by Eq.~\ref{equ:QuadKin}.

Following Ref. \cite{Dod97},
we find the foreground degradation factor (FDF) as
\begin{equation}
{\rm FDF} = \left[\frac{1}{1-(g^0 \cdot g^1)^2/g^1 \cdot g^1} \right]^{1/2}\, ,
\end{equation}
where the dot product is defined as
\begin{equation}
A \cdot B \equiv \sigma_{{\cal P}_x}^{(0)2} \sum_{a,b=1}^{N_{\rm ch}} A_a C_{ab}^{-1} B_b\, ,
\end{equation} 
with variance in the absence of the kinematic quadrupole contribution given by
\begin{equation}
\sigma_{{\cal P}_x}^{(0)2} = \frac{1}{\sum_{a=1}^{N_{\rm ch}}
\left[C_{ab}^x\right]^{-1}}\, .
\end{equation}
Finally, the variance in the presence of the  kinematic quadrupole
contribution is given by
\begin{equation}
\sigma_{{\cal P}_x}^2 = ({\rm FDF})^2 \sigma_{{\cal P}_x}^{(0)2} \, .
\end{equation}

\begin{figure}[!h]
\centerline{\psfig{file=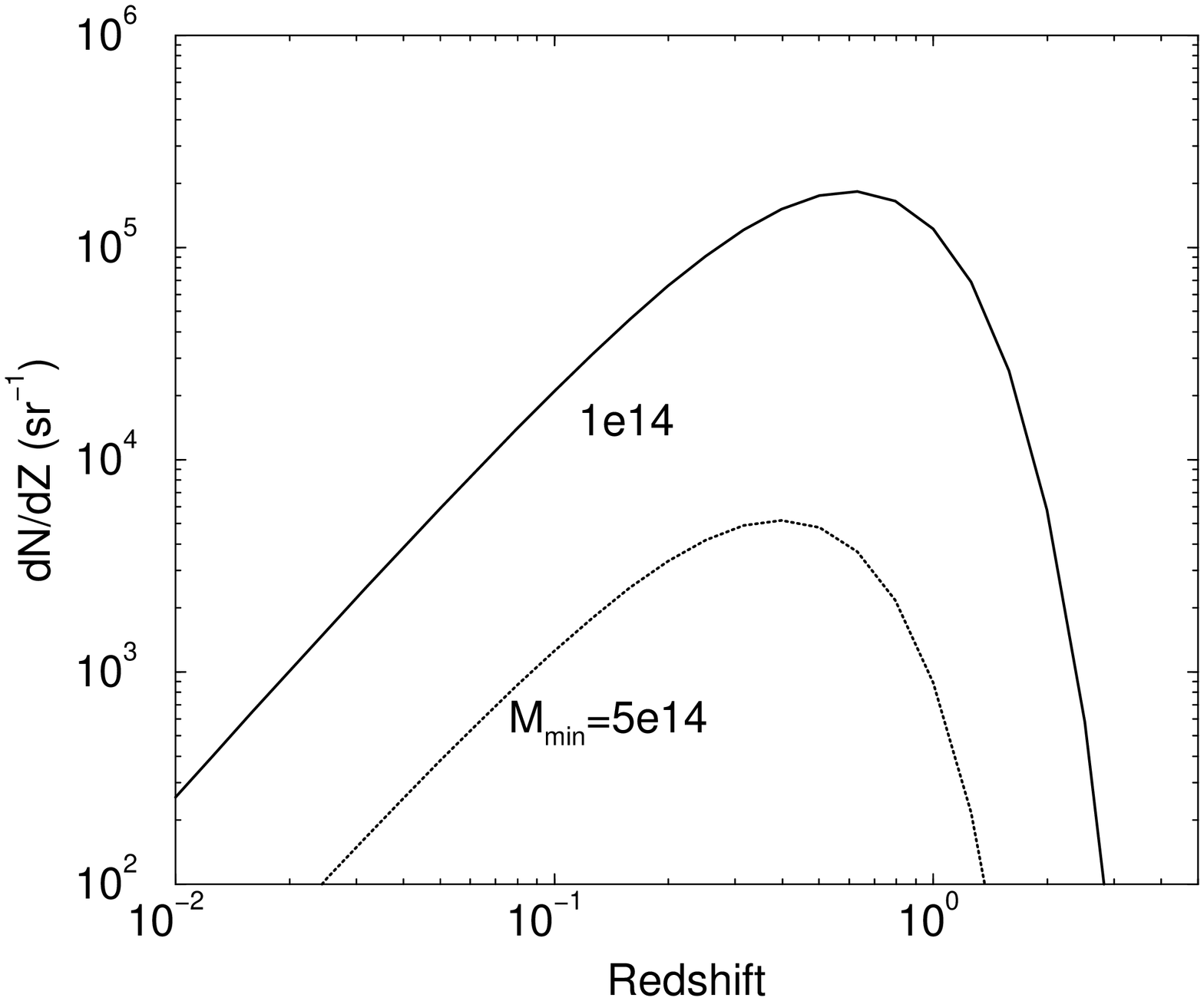,width=3.5in,angle=-90}}
\centerline{\psfig{file=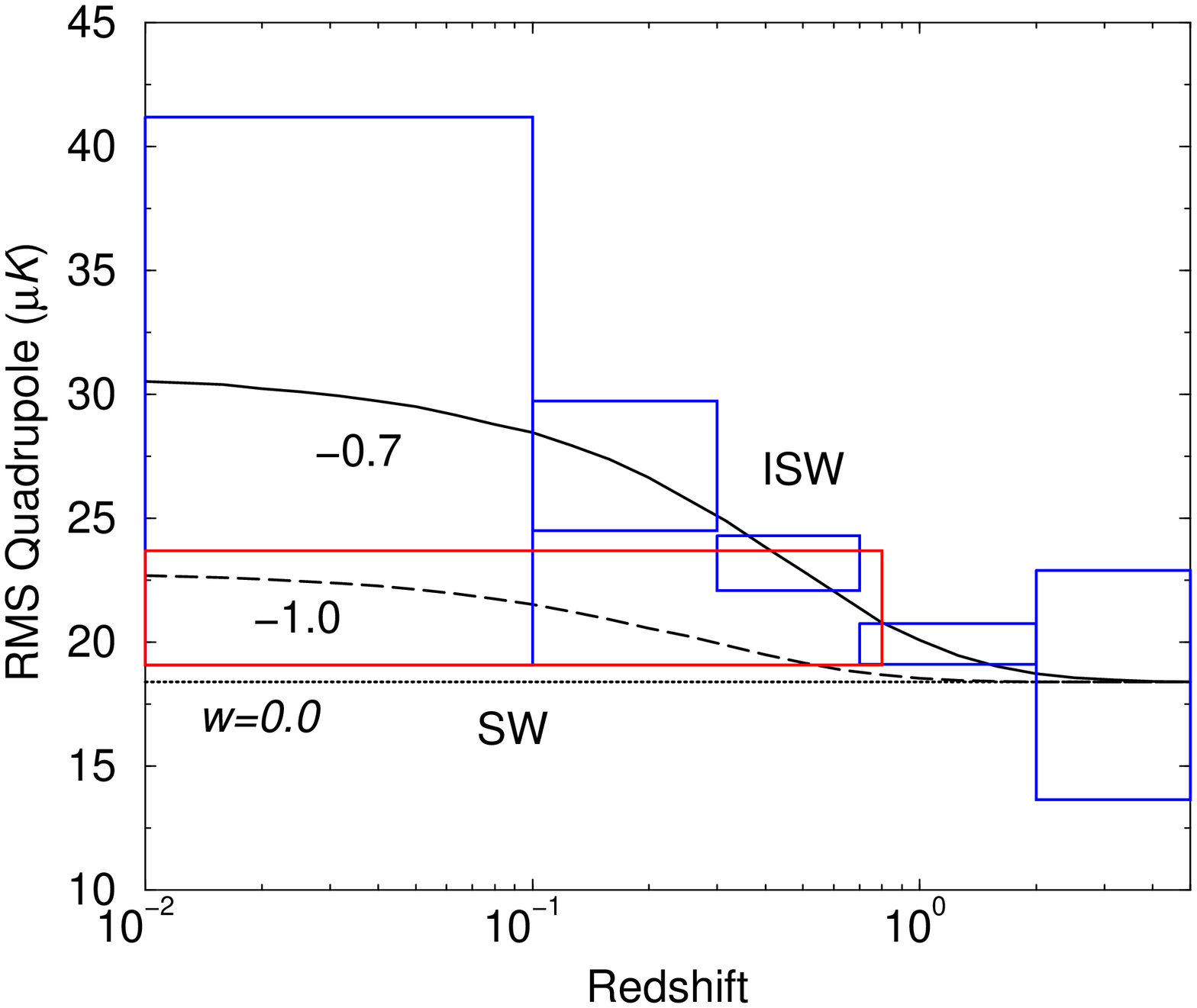,width=3.5in,angle=-90}}
\caption{{\it Top:} The expected redshift distribution of galaxy clusters in
SZ surveys. The upcoming surveys, such as the one planned 4000 deg$^2$
observations with the South
Pole Telescope, can probe mass limits down to 10$^{14}$ M$_{\sun}$ and with Planck surveyor
down to $5 \times 10^{14}$ M$_{\sun}$.
{\it Bottom:} The rms temperature 
quadrupole as a function of redshift. The error
bars on the top curve illustrate 
the expected uncertainties in the reconstruction
of rms temperature quadrupole from followup multi-frequency polarization
observations of galaxy clusters with a catalog down to  10$^{14}$ M$_{\sun}$ and covering
10,000 deg$^2$ (see text for details). The large single
error on the middle curve illustrate the expected uncertainty with the whole-sky
Planck cluster catalog down to a mass limit of $5 \times 10^{14}$ M$_{\sun}$ 
and using Planck polarization observations of same clusters.}
\label{fig:qrms}
\end{figure}

\subsection{Primordial Quadrupole Reconstruction}
We can calculate the expected errors in the primordial quadrupole
reconstruction by considering the enhancement in the instrumental noise
due to the separation of the contaminant kinematic quadrupole contribution.
Under the assumption that the spectral dependence of the kinematic
quadrupole, $g(x)$, is known, we can write the final variance
of the determined quadrupole polarization as
\begin{equation}\label{equ:error}
\sigma^2_{\cal{P}} = 2
({\rm FDF})^2 \left(\sum_{i=1}^{N_{\rm ch}} 1/\sigma_i^2\right)^{-1} \, .
\end{equation}
Here, for simplicity, we have taken the
instrumental noise covariance matrix to be diagonal with variance in each 
of the polarization components given by $\sigma_i^2$.  The factor of 2 
accounts for the two polarization components, $(P_x, P_y)$,
with equal variance. For the case of equal variance in all channels, 
$\sigma =\sigma_i$, $\sigma^2 \rightarrow 2 ({\rm FDF})^2 
\sigma^2/N_{\rm ch}$ \cite{Dod97}. 

Given the uncertainty in the primordial polarization determination, Eq.~\ref{equ:error},
we can calculate the expected uncertainty in the rms quadrupole
by considering Eq.~1.  In Fig.~2, we show the redshift
distribution of clusters expected from planned SZ surveys as 
calculated using the Press-Schechter \cite{PreSch74} mass function
based on analytic approaches in the literature \cite{CooShe02}.
Note that the selection functions of clusters in SZ surveys are
essentially constant in mass over a wide range in redshift.
An experiment such as the planned South Pole Telescope is 
likely to detect clusters down to a mass of 10$^{14}$ M$_{\sun}$ 
\cite{Haietal00}.
Here, we assume a similar survey but roughly a factor of 2.5 increase
in sky coverage with a total area of $\sim 10,000$ deg$^2$.

We bin the redshift distribution of these clusters in five bins between
redshifts of 0 and 5. We calculate the optical depth to scattering in
each of these clusters by assuming constant gas mass fractions
and converting the gas mass in each cluster to a number density of
electrons under spherical arguments. These optical
depths range from 0.005 to 0.02 and typical polarization  contributions, 
due to the temperature quadrupole,
range from 0.02 to 0.1 $\mu$K. While 
one would require high sensitive observations with an instrumental 
noise contribution below this level to detect the polarization
contribution reliably towards an individual cluster,
for the study suggested here, this is not 
necessary. Since we are attempting
to make a statistical detection of the rms
quadrupole, one can average over large samples of clusters
widely distributed on the sky and the final variance on the rms
quadrupole can be improved, in principle, by the number of
individual clusters.

To consider the frequency separation of the cluster polarization
contributions, we assume an experiment with observational frequencies
at 30, 100, 200 and 400 GHz with a final polarization sensitivity of
1 $\mu$K in each of the channels. 
The frequency degradation factor (FDF) in this case is 2.07. 
For polarization sensitive Planck channels, 
at 147, 217 and 545 and published sensitivities ranging from 10 to 560 $\mu$K,
we determined a FDF of 4.1. In general a low frequency channel 
at the Rayleigh-Jeans
limit is required to reduce the degradation in noise, due to
frequency cleaning, to factors of 2 and below.

Using the FDF factor for our four channel experiment and
the redshift distribution of clusters, we
can now determine how well the rms quadrupole can be established.
We illustrate the expected rms quadrupole and errors around the top
curve in Fig.~2.  Here, we have taken a flat cosmological model with a
matter density, in units of the critical density,
$\Omega_m=0.35$, baryon density $\Omega_b=0.05$, dark-energy
density $\Omega_x=0.65$ with an equation state $w=-0.7$, 
a Hubble constant $h=0.65$, and spectral index of $n=1$ for
primordial density perturbations.   We have normalized the fluctuations to
COBE \cite{BunWhi97}. For illustration purposes, we also show the 
expected rms temperature quadrupole for $w=0$ and $w=-1$;
the former corresponds to the case where dark-energy behaves as matter
while the latter is the usual cosmological constant case.

We also make use of the Planck cluster catalog as well as Planck
polarization observations with expected sensitivities. The Planck
SZ cluster catalog  is expected to detect clusters down to a mass limit
of $5 \times 10^{14}$ M$_{\sun}$ \cite{Kayetal01}. 
While the number counts per steradian
is lowered (Fig.~2), one gains by the fact that the Planck catalog is
all-sky. Still, due to poor sensitivity of polarization 
observations, the reconstruction is noise dominated and no useful information can be obtained by binning
the catalog as a function of redshift. Thus, we combine all data to a single
bin to estimate the rms temperature quadrupole. With Planck,
one can determine the excess quadrupole, but it is unlikely that
detailed information on its evolution can be established. We therefore
suggest followup observations of the same cluster catalog either through individual
targeted observations or as part of a post-Planck polarization mission
that may attempt to detect the signature of inflationary gravitational waves.

In estimating quadrupole reconstruction errors, we have ignored uncertainties associated with
the determination of optical depth in each cluster. In the case of Planck data, 
if the optical depth can be established at the level of few tens of percent or better,
the final uncertainty in the temperature quadrupole reconstruction from polarization
will be dominated by noise associated with polarization measurements and not in the determination of
individual cluster optical
depth; Given current expectations based on cluster scaling relations, it is very likely that
one cen determine the optical depth to tens of percent or better in upcoming data.

For the suggested improved polarization experiment, the situation, however is different.
For the polarization noise there to dominate the final quadrupole reconstruction, optical depths to
individual clusters should be known at the level of few percent or better. Such a determination
will require an improved understanding of cluster gas physics so that one can reliably convert the
observed SZ thermal and kinetic temperature decrements to measurements of optical depth.
During the Planck era, with the availability of multifrequency data, we
expect such a determination may be possible. Furthermore, potential studies with clusters, such as the one
proposed here, motivate improvements to our understanding of cluster physics and we do not consider
our inability to determine optical depths to be the limitating factor.

\section{Discussion}
\label{sec:results}

As illustrated  in Fig.~2, a CMB temperature and polarization 
survey of 10,000 deg$^2$ down to a fixed mass
limit of 10$^{14}$ M$_{\sun}$ can be used to reconstruct the rms temperature 
quadrupole and its evolution reliably. The temperature information is used
to detect and catalog clusters as well as to determine the optical
depth to scattering. The same cluster sample is then followed up with
multi-frequency polarization observations. In reconstructing the
primordial temperature quadrupole, one is using these 
clusters as a tracer population in which each cluster individually 
provides information on the  quadrupole. 

The first attempt to perform a study similar to this will come from
Planck's temperature and polarization observations. While Planck
will provide an adequate size catalog of clusters distributed over
the whole sky, the poor sensitivity of polarization observations, however,
limits one from extracting detailed information on the evolution of the
temperature quadrupole. The Planck catalog, however, can be used to potentially detect the
excess  of the temperature quadrupole rms at low redshifts due to the ISW effect.

For a more detailed detection of the ISW effect, including its redshift evolution,
one needs an order of magnitude higher sensitive polarization observations than Planck. 
An opportunity for such a study can be considered with respect to planned SZ 
surveys with telescopes such as the South Pole Telescope (SPT). If these upcoming observations can be made 
sensitive to polarization, the cluster catalog detectable from such wide field SZ surveys can
potentially be used for a more reliable reconstruction of the ISW signature. The advent of
polarization sensitive bolometers suggest that such possibilities are within reach over the next decade.

The suggested reconstruction here is beyond what was considered in Ref. \cite{KamLoe97}. Here, in addition to
establishing the quadrupole, we have suggested the use of its evolution
to directly observe the presence of the ISW effect. Since 
large angular scale temperature observations or cross-correlations of the
temperature with maps of the local universe cannot be used to establish the
ISW effect reliably, cluster polarization  observations provide 
an indirect method for this purpose. 

The reconstruction, however, is subjected to several confusions.
Here, we have considered the confusion associated with the kinematic quadrupole
contribution and have suggested the use of multi-frequency
polarization observations as a way to separate the two contributions.
With adequate frequency coverage, this separation can be achieved with
an increase in the variance  of 
polarization noise by a factor of up to  $\sim$ 2.

In addition to the kinematic quadrupole polarization, galaxy clusters also
produce polarization signals associated with two other mechanisms that
generate a local quadrupole. These effects are due to the intrinsic quadrupole
associated with double scattering effects within a single
cluster. After the first scattering, the CMB acquires an local
anisotropy both due to thermal and kinematic effects. Second
scattering then produces polarization. The polarization
associated with the thermal effect is proportional to
$\tau^2kT_e/m_ec^2$. The kinematic double scattering induces a
polarization signal proportional to $\tau^2\beta$. The frequency
dependence of the polarization will be the same as that of the thermal
or kinematic effect, respectively.
Both second scattering effects  are expected to be small due to
their dependence on $\tau^2$. 
Additionally, for double scattering with a kinematic intrinsic
quadrupole, 
when averaging over large samples, one expects $\beta\approx 0$
due to the randomness associated with the large scale velocity field.
This will further suppress the contaminant signal. 
Including the double scattering associated with a thermal intrinsic
quadrupole leads to biases in the rms quadrupole estimate of the order of a few
percent in $\mu$K.
For unbiased estimates of the quadrupole,
one can potentially account for this contaminant
with prior knowledge on $\tau$ and $T_e$.

Besides establishing the presence of the ISW effect reliably, the
reconstruction is also useful for cosmological purposes. As illustrated in
Fig.~2, the ISW contribution, and thus the rms quadrupole, is
strongly sensitive to basic properties of the dark energy such as its
equation of state. Though not considered in this paper, the ISW contribution also
varies significantly with clustering aspects of the dark energy component
beyond the usual smooth assumption, such as in generalized dark matter models
of Ref. \cite{Hu98}. Considering a Fisher-matrix approach to the
measurements shown in Fig.~2, assuming a cosmological model
with the fiducial parameters listed earlier, we determined that one can extract the
equation of state of the dark energy, $w$ with an uncertainty of $\sim$ 0.1.
Beyond estimates of cosmology, we suggest that in the future
a reconstruction such as the one discussed here should be considered to
establish  the presence of the ISW effect and its evolution.

\acknowledgments
This work was supported in part by DoE DE-FG03-92-ER40701
a senior research fellowship from the Sherman Fairchild foundation (AC), and
a summer undergraduate research fellowship from Caltech (DB).
AC thanks the Kavli Institute for Theoretical Physics 
(supported by NSF PHY99-07949) for the hospitality while this work
was completed. DB thanks the Astrophysics Group of the Cavendish
Laboratory where parts of this work were carried out. DB especially thanks
Anthony Challinor for interesting comments on the ``quadrupole tomography''.

\end{document}